\documentclass[10pt,twocolumn]{article}

\usepackage{amsmath}
\usepackage{multirow}
\usepackage{graphicx}
\usepackage{caption}
\usepackage[margin=2cm]{geometry}
\usepackage{etoolbox}
\usepackage{subfig}
\usepackage{adjustbox}
\usepackage{array}
\newcolumntype{C}[1]{>{\centering}m{#1}}
\usepackage{authblk}
\usepackage{upgreek}
\usepackage{xfrac}
\usepackage{booktabs}
\usepackage{color}
\usepackage{ulem}
\usepackage[utf8]{inputenc}

\title{Critical pressure asymmetry in the enclosed fluid diode}
\author[1]{Jack R. Panter}
\author[2]{Yonas Gizaw}
\author[1]{Halim Kusumaatmaja}
\affil[1]{Department of Physics, Durham University, South Road, Durham, DH1 3LE}
\affil[2]{The Procter and Gamble Co., Mason Business Center, 8700 S. Mason-Montgomery Road, Mason, OH, USA}

\date{}
\begin{document}

\twocolumn[
\begin{@twocolumnfalse}
\maketitle

\begin{abstract}

Joint physically and chemically pattered surfaces can provide efficient and passive manipulation of fluid flow. The ability of many of these surfaces to allow only unidirectional flow mean they are often referred to as fluid diodes. Synthetic analogues of these are enabling technologies from sustainable water collection via fog harvesting, to improved wound dressings. One key fluid diode geometry features a pore sandwiched between two absorbent substrates, an important design for applications which require liquid capture while preventing back-flow. However, the enclosed pore is particularly challenging to design as an effective fluid diode, due to the need for both a low Laplace pressure for liquid entering the pore, and a high Laplace pressure to liquid leaving. Here, we calculate the Laplace pressure for fluid travelling in both directions on a range of conical pore designs with a chemical gradient. We show that this chemical gradient is in general required to achieve the largest critical pressure differences between incoming and outgoing liquids. Finally, we discuss the optimisation strategy to maximise this critical pressure asymmetry.

\end{abstract}


\vspace{5mm}

\end{@twocolumnfalse}
]

\section*{Introduction}
Structured surfaces which control the direction of motion of liquid droplets are prevalent in nature \cite{Li2019}. Strong directionality is enabled by surfaces which have both physical and chemical gradients, demonstrated for example by the textured conical spines of the cactus \textit{Opuntia microdasys} \cite{Ju2012}, the spindle-knots of spider silks \cite{Bai2010}, and the ratcheted surface of butterfly wings \cite{Kusumaatmaja2009}. This directionality is a result of a driving force from the combined effects of a Laplace pressure gradient across the droplet, caused by the physical structure \cite{Renvoise2009,Lorenceau2004}, and a surface energy gradient under the droplet, caused by a chemical pattern \cite{Brochard1989, Gennes2010}. 

Inspired by these biological examples, there is substantial interest in synthesising structures which enforce unidirectional liquid flow - fluid diodes \cite{Mates2014}. The technological applications of fluid diodes span numerous and ambitious fields focussing on efficiency and sustainability \cite{Zhang2019}, such as oil-water separation \cite{Zhao2017} and water purification or fog harvesting \cite{Brown2016}. Fluid diodes are being realised in a range of geometries, such as across surface structures \cite{Li2017}, along a porous strip \cite{Shou2018}, through the thickness of a material \cite{Zhao2017}, and within microfluidic channels \cite{Zimmermann2008}. 

\begin{figure}[!b]
\includegraphics[width=0.5\textwidth]{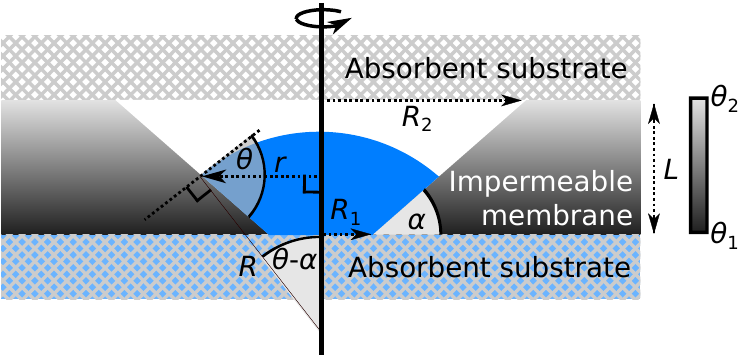}
\caption{2D illustration of the axisymmetric pore construction and outgoing meniscus profile, the axis of symmetry shown as the thick vertical line. Liquid is shown in blue, with vapour shown in white.}
\label{fig:outsetup}
\end{figure}

The optimal performance of a fluid diode relies on maintaining a high contrast in the force required to transport fluid through the diode in the forward direction, compared to the reverse direction. One geometry in which this remains particularly challenging is the enclosed pore, illustrated in Fig. \ref{fig:outsetup}. In this geometry, a pore through an impermeable membrane is sandwiched between two absorbent substrates. The diode ability here arises from the critical pressure asymmetry - the difference in the maximum Laplace pressure (critical pressure) required to force liquid from the bottom substrate to the top substrate, compared to the reverse direction. Such a design is particularly suited to a range of applications in which fluid should be readily absorbed into the diode, but not be able to pass back out. Cleaning and hygiene are two notable areas where such applications are prominent. In these, the diode would both facilitate absorption of liquid from a surface, such as skin, into a porous material while also preventing back-flow out of the material. General and widely-used potential applications include diapers, cloths, and towels \cite{Diersch2010}. However, the fluid diode is also gaining interest in high-performance innovations, such as sports textiles which absorb and remove sweat to cool the body, but are waterproof from the outside \cite{Miao2018}; and wound dressings, in which excess fluid should be selectively absorbed out of the wound to improve healing and reduce infection risk \cite{Shi2018}.

Here, we explore the diode capabilities of a conical pore augmented with a chemical gradient. Conical pores, or pores with a variation in cross-sectional width have been shown in microfluidic fields to enable effective passive regulation of fluid flow, with a key application being the capillary burst valve \cite{Zimmermann2008,Cho2007, Chen2008,Taher2018}. Furthermore, substantial progress has been made in calculating the maximum Laplace pressures for liquid entering physically textured surfaces, see for example \cite{Kaufman2017, Panter2019}, as well as liquid exiting physically textured surfaces of axisymmetric and non-axisymmetric cross sections \cite{Ma2019,Agonafer2018}. However, the enclosed geometry, efficacy at preventing back-flow, and the impact of chemical patterning have never been discussed.

In the Outgoing critical pressures section, we begin by calculating the Laplace pressure for liquid leaving the pore. In the Incoming critical pressures section, we calculate the Laplace pressure for liquid entering the pore. We then compare the incoming and outgoing maximum Laplace pressures, using the critical pressure asymmetry to measure the strength of the diode in the Critical pressure asymmetry section, before finally optimising the chemical pattern to produce the maximum possible critical pressure asymmetry in the Optimum asymmetry section.

\section*{Results and discussion}
\subsection*{Outgoing critical pressures}
\label{sec:out}
\subsubsection*{Setup}

The model setup, illustrated in Fig. \ref{fig:outsetup}, features a liquid-impermeable membrane shown in shaded grey, punctured by an axisymmetric (conical) pore of wedge angle $\alpha$. Without loss of generality, we restrict $\alpha$ to the interval $[0,\pi/2]$, so that the smallest pore radius $R_1$ is always located at the bottom of the system, and the largest pore radius $R_2$ is located at the top. For $\alpha > \pi/2$, we need not perform additional calculations, but rather turn the pore as shown upside-down, and exchange the roles of incoming and outgoing critical pressure. In addition to a physical gradient, we employ a chemical gradient in the form of the local contact angle $\theta(r)$ which varies from $\theta_1$ at the bottom of the pore to $\theta_2$ at the top. Although any variation in contact angle can be chosen, we employ a linear variation to most closely compare with the linear physical gradient of the conic profile, where
\begin{equation}
\theta(r) = \theta_1 + (\theta_2-\theta_1)\frac{r-R_1}{R_2-R_1}.
\end{equation}
As the primary focus of this section is to model the pressure required for liquid to exit the pore, the contact angles used in the analysis throughout should be treated as the advancing contact angles on a surface where hysteresis is present.
   
In a fully enclosed pore, the top of bottom surface of the liquid-impermeable membrane are in contact with liquid-absorbent substrates, shown as cross-hatched areas in Fig. \ref{fig:outsetup}. For considering the outgoing critical pressure, the bottom absorbent substrate is modelled as an infinite liquid reservoir from which the liquid meniscus rises upwards into the pore. The top substrate is modelled as a perfect liquid sink: as soon as liquid reaches the top of the pore or contacts the upper surface, the diode breaks down.

We consider the surface to be smooth with the only pinning sites occurring at the sharp corners at the top and bottom of the pore, and we work below the capillary length so that the liquid meniscus assumes a spherical cap geometry for all values of the contact line radius $r$. To ensure this, the pore size should typically be less than several millimetres, as for example, the capillary length of water is 2.7 mm, while for a low surface tension liquid such as hexane, the capillary length is 1.7 mm. We also consider the system to be larger than the longest range van der Waals forces (approximately 100 nm \cite{DeGennes2004}), so that disjoining-pressure modifications to the liquid-vapour interface shape close to the contact line are negligible. The pressure difference $\Delta P$ across the liquid-vapour interface is therefore described by the Young-Laplace equation appropriate for a spherical geometry: $\Delta P = 2\gamma_{\rm{lv}} / R$, where $\gamma_{\rm{lv}}$ is the liquid-vapour interfacial tension and $R$ is the radius of the sphere. This spherical cap model also implies we treat the fluid configurations as static; the impact of fluid velocity on burst pressures can also be important, but is outside the scope of the current work. For instance, such dynamical effects have been studied in a variety of porous structures \cite{Agonafer2018,Moebius2012,Rabbani2019}. Throughout, we nondimensionalise the Laplace pressure so that $\Delta P_r =  \Delta P / ( 2\gamma_{\rm{lv}} / R_1)$. For convenience, we also nondimensionalise all radii with respect to $R_1$, so that for example, $R' = R/R_1$ and $R_2'=R_2/R_1$.

We note here that although we label the fluids as 'liquid' and 'vapour', as the methods only require a knowledge of the contact angle at the three-phase contact line and the fluid-fluid interfacial tension, the analyses presented here are entirely general for any pair of immiscible fluids, such as oil and water.

Using the construction in Fig. \ref{fig:outsetup}, the outgoing Laplace pressure $\Delta P^{\rm{out}}_r$ may be described as function of the reduced contact line radius $r'=r/R_1$,
\begin{equation}
\Delta P^{\rm{out}}_r = \frac{1}{r'} \sin \left[ \theta_1 + (\theta_2-\theta_1)\frac{r'-1}{R_2'-1} - \alpha \right]. \label{eqn:DELTA_P_OUT}
\end{equation}

\begin{figure}[!t]
\includegraphics[width=0.5\textwidth]{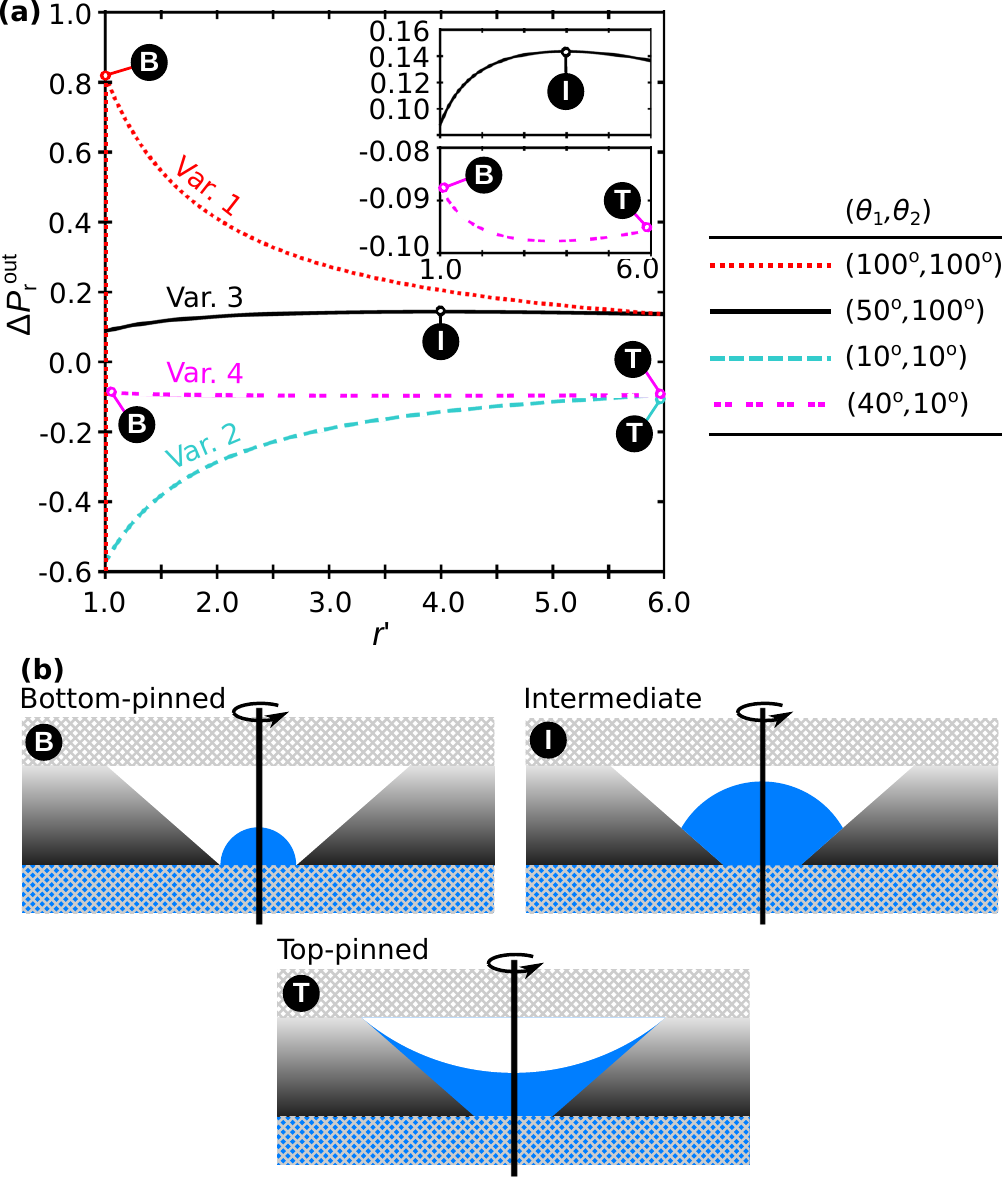}
\caption{(a) Example plots of each of the four outgoing reduced pressure variations with $r'$, with insets magnifying the local maxima/minima. The local maxima of each variation is associated with one of three critical meniscus types, illustrated in (b).}
\label{fig:outtypes}
\end{figure}
Finding the critical outgoing pressure $\Delta P^{\rm{out}}_c$ then becomes finding the maximum value of $\Delta P_r$ as $r'$ is increased from $1$ to $R_2'$. As soon as the contact radius reaches $R_2'$, the liquid will be spontaneously absorbed into the top substrate. It is possible that the apex of the meniscus contacts the upper substrate before $r'=R_2'$, but we reserve discussion of these cases to the section entitled "Influence of the top substrate: B' and I' critical menisci". To begin with, if we only allow liquid to be absorbed into the top substrate at $r'=R_2'$, then $\Delta P^{\rm{out}}_r$ exhibits four characteristic variations with $r'$, depending on $R_2'$, $\theta_1$, $\theta_2$, and $\alpha$; representative examples of each are plotted in Fig. \ref{fig:outtypes}(a), where we fix $\alpha=45^\circ$, $R_2'$=6. 

\subsubsection*{Variation 1 and the bottom-pinned (B) critical meniscus}
The first variation, shown as the dotted red line, shows that for $r' > $  1, $\Delta P^{\rm{out}}_r$ decreases monotonically with $r'$. When $r'$ = 1 however, the contact line is pinned to the bottom of the pore. The Gibbs pinning criterion of a contact line at a sharp corner \cite{Gibbs1906,Blow2009} then permits a continuum of allowed pressures, as the pinned contact angle may vary from $\theta_1$ with respect to the bottom surface of the impenetrable membrane, to $\theta_1$ with respect to the sloping pore wall. This is shown as the vertical dotted red line at $r'$ = 1. The critical pressure here occurs in the bottom-pinned state, labelled the B-state in Fig. \ref{fig:outtypes}(b), where,
\begin{equation}
\Delta P^{\rm{out}}_c(\rm{B})=\sin\left[ \min(\theta_1-\alpha, \frac{\pi}{2}) \right]. \label{eqn:P_B}
\end{equation}
For $\theta_1-\alpha < \pi/2$, $\Delta P^{\rm{out}}_c(\rm{B})$ occurs when the interface depins from the sharp corner, such that the contact angle at the contact line is equal to $\theta_1$. For $\theta_1-\alpha > \pi/2$ however, $\Delta P^{\rm{out}}_c(\rm{B})$ happens when the contact angle reaches $\pi/2 - \alpha$, before the depinning event, because the maximum possible critical pressure for the system is attained here at $R=R_1$.
 
\subsubsection*{Variation 2 and the top-pinned (T) critical meniscus} 
The second variation, shown as the dashed cyan line in Fig. \ref{fig:outtypes}(a), shows a monotonic increase of $\Delta P^{\rm{out}}_r$ with $r'$. The critical pressure therefore occurs at the point when the contact line reaches the top of the system at $r'=R_2'$, where $\theta = \theta_2$. This is labelled the top-pinned (T) state in Fig. \ref{fig:outtypes}(b). In this case,
\begin{equation}
\Delta P^{\rm{out}}_c(\rm{T})= \frac{1}{\mathit{R}_2'}\sin\left( \theta_2 - \alpha \right). \label{eqn:P_T}
\end{equation}

\subsubsection*{Variation 3 and the intermediate (I) critical meniscus}
The third variation, shown as the solid black line in Fig. \ref{fig:outtypes}(a), exhibits a non-monotonic variation with $r'$, and a local maximum at intermediate values of  $r'$, labelled the I-state in Fig. \ref{fig:outtypes}(b). The upper inset panel highlights the local maximum in a vertical magnification. To solve for the critical pressure, we aim to find stationary points of $\Delta P^{\rm{out}}_r$ in Eq. \eqref{eqn:DELTA_P_OUT}, such that the critical contact line radius $r_c' \in (1, R_2')$. This amounts to solving,
\begin{align}
&\frac{1}{r_c'}\left(\frac{\theta_2-\theta_1}{R_2'-1}\right)\cos\left(\theta_1+(\theta_2-\theta_1)\frac{r_c'-1}{R_2'-1}-\alpha\right) \nonumber \\ 
 -&\frac{1}{r_c'^2}\sin\left(\theta_1+(\theta_2-\theta_1)\frac{r_c'-1}{R_2'-1}-\alpha\right) = 0, \label{eqn:P_I}
\end{align}
for $r_c'$, such that $\Delta P^{\rm{out}}_r$ is maximal, yielding $\Delta P^{\rm{out}}_c(\rm{I})$. In general, this is not analytically solvable and instead must be solved numerically. Interestingly, such a local maximum cannot exist for a chemically homogeneous pore: rather it is result of the competition between physical and chemical gradients. To illustrate this, we consider the example shown in Fig. \ref{fig:outtypes}(a) (solid black line), for which $\theta_1 < \pi/2$, but $\theta_2 > \pi/2$. Physically, as the contact line radius $r'$ increases from 1 to $R_2'$, this tends to increase the droplet radius $R'$ and hence reduce the magnitude of the Laplace pressure. Chemically, the simultaneous increase in local contact angle tends to reduce the droplet radius $R$ and so increase the droplet pressure. When the I-state exists, it is therefore due to the balancing of these two effects.

\subsubsection*{Variation 4 and the B and T critical menisci}
Instead of a local maximum, the fourth variation, shown as the double-dashed magenta line in Fig. \ref{fig:outtypes}(a), exhibits a local minimum. The lower inset panel highlights the local minimum in a vertical magnification. This behaviour is observed when solving Eq. \eqref{eqn:P_I} which yields a minimal solution of $\Delta P^{\rm{out}}_r$. Thus, both the B-state at $r'=1$ and the T-state at $r'=R_2'$ become local maximisers of $\Delta P^{\rm{out}}_r$. Which state globally maximises $\Delta P^{\rm{out}}_r$ is found by comparing Eq. \eqref{eqn:P_B} and Eq. \eqref{eqn:P_T}. We detail this comparison in the Outgoing critical pressures visualisation section. We further note that here, both the B-state and T-state have negative Laplace pressures. It is possible for the B-state to have negative Laplace pressure (whereby $\theta_1 < \alpha$) if $\theta_2$ is so small that the Laplace pressure becomes more negative on increasing $r'$ from 1 to $R_2'$,

\subsubsection*{Influence of top substrate: B' and I' critical menisci}
\begin{figure}[!t]
\includegraphics[width=0.5\textwidth]{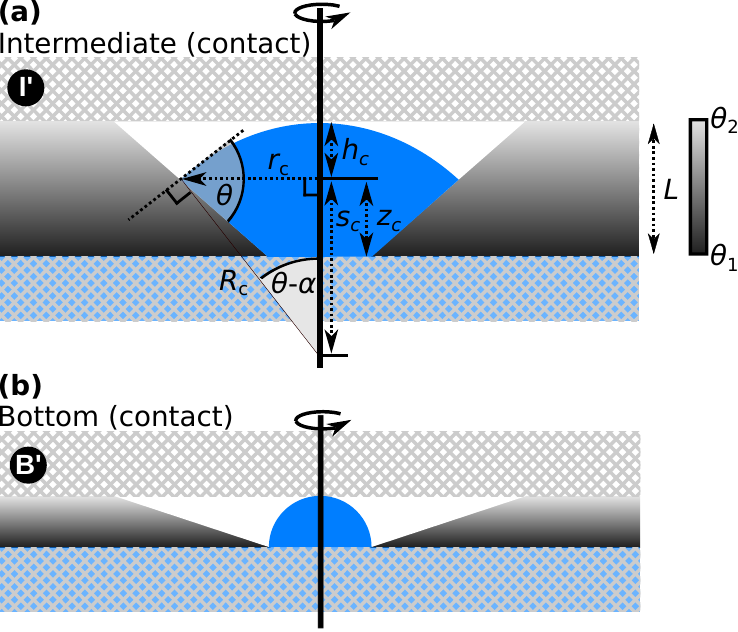}
\caption{(a) Construction used to calculate the critical pressure of the I' state. (b) Illustration of the B' state, with the point of failure highlighted by a red circle.}
\label{fig:outtypesprime}
\end{figure}

When the liquid meniscus is convex, the centre of meniscus may contact the top of the pore before the B-state or I-state critical pressure is reached. We denote the bottom-pinned contacting state B', and the intermediate contacting state I'. We note that a top-pinned contacting state cannot occur, as this would require the centre of the meniscus to contact the top absorbent substrate before the three-phase contact line. For clarity of notation throughout, we refer to a liquid meniscus as being convex if the droplet forms a converging lens, such as the B-state in Fig. \ref{fig:outtypes}(b), and concave if the droplet forms a diverging lens, such as the T-state in Fig. \ref{fig:outtypes}(b). In Fig. \ref{fig:outtypesprime}(a), we construct the total height of the liquid meniscus as the sum of the height of the contact line $z_c$ above the pore bottom, and the height of the spherical cap above this $h_c$. Noticing that $h_c = R_c - s_c$, where $s_c$ is the $z$-distance from the centre of the spherical cap to the contact line, we derive,
\begin{equation}
h_c=r_c\frac{1-\cos(\theta(r_c)-\alpha)}{\sin(\theta(r_c)-\alpha)}.
\end{equation}
For the spherical cap to touch the upper substrate, $z_c+h_c=L$ must be satisfied, where $z_c = \left(r_c-R_1\right)\tan\alpha$, and the membrane thickness $L = \left(R_2-R_1\right)\tan\alpha$. In reduced units, this amounts to solving,
\begin{equation}
\left(r_c'-R_2'\right)\tan\alpha + r_c'\frac{1-\cos(\theta(r_c')-\alpha)}{\sin(\theta(r_c')-\alpha)}=0.
\end{equation}
In general this does not have analytic solutions and must be solved numerically. Once $r_c'$ is found in this way, it is straightforward to substitute $r'$ for $r_c'$ in Eq. \eqref{eqn:DELTA_P_OUT} to recover the critical pressure $\Delta P^{\rm{out}}_c(\rm{I'})$ caused by the cap contacting the upper substrate, while the contact line radius takes an intermediate value between $R_1$ and $R_2$. 

If instead the contact line is pinned to the bottom of the pore at the point of meniscus contact, as illustrated in Fig. \ref{fig:outtypesprime}(b), the B'-type critical meniscus arises, where the outgoing critical pressure may be simply expressed as,
\begin{equation}
\Delta P^{\rm{out}}_c(\rm{B'})= \frac{2}{\mathit{L}'+\frac{1}{\mathit{L}'}}, \label{eqn:P_BPRIME}
\end{equation} 
where $L'=L/R_1$.

\subsubsection*{Critical morphology existence ranges}
Overall, five different critical interface morphologies may occur: B, B', T, I, and I', in which the associated critical pressures feature different dependencies on $\theta_1$, $\theta_2$, $\alpha$, and $R_2'$. Despite this complexity, the system parameters can be partitioned into four categories, determined based on whether the liquid meniscus is convex ($\theta > \alpha$) or concave ($\theta < \alpha$) at the top and bottom of the system. In Table \ref{table:outtypes}, we show which critical morphology is possible within each category.
\begin{table}[!h]
\small
\caption{The critical outgoing meniscus types able to occur for a convex meniscus, $\theta > \alpha$, or concave meniscus $\theta < \alpha$.}
\label{table:outtypes}
	\begin{tabular*}{0.48\textwidth}{@{\extracolsep{\fill}}lll}
	\hline
	& $\theta_1>\alpha$ & $\theta_1<\alpha$ \\
	\hline
	$\theta_2>\alpha$ & B, B', I, I' & I, I' \\
	$\theta_2<\alpha$ & B, B' & B, T \\
	\hline
	\end{tabular*}
\end{table}

For ($\theta_1 > \alpha$, $\theta_2 < \alpha$), the critical meniscus must occur when contact line is pinned to the bottom of the system in B or B'. For ($\theta_1 < \alpha$, $\theta_2 > \alpha$) however, the Lapalce pressure is negative when the contact line is at the bottom of the pore, and positive at the top, so the critical meniscus must occur in some intermediate state: I or I'. For ($\theta_1 > \alpha$, $\theta_2 > \alpha$), the meniscus is convex for all $r$, meaning the B, B', I, or I' states could occur. For ($\theta_1 < \alpha$, $\theta_2 < \alpha$), the meniscus is concave for all $r$, so that the critical pressure must either occur at the bottom of the system, as B, or top, as T.

\subsubsection*{Outgoing critical pressures visualisation}
\label{sec:out_vis}
\begin{figure*}[!ht]
\includegraphics[width=\textwidth]{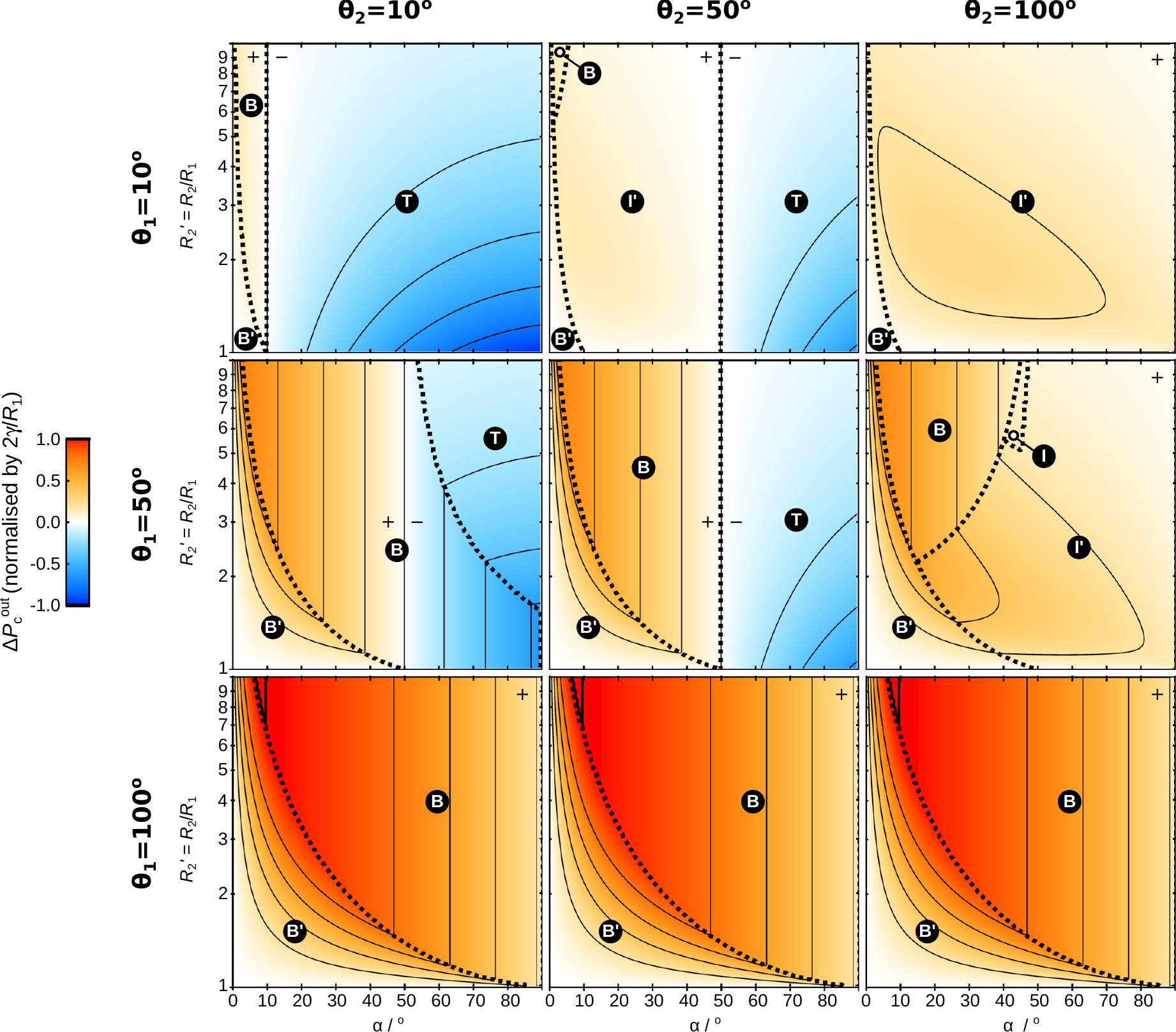}
\caption{Matrix of $R_2'$ - $\alpha$ contour plots of the outgoing critical pressure for a selection of $\theta_1$ and $\theta_2$. The outgoing meniscus types are labelled with black circles. The boundaries between these critical types are shown as dotted black lines. Contours are shown at intervals in $\Delta P_c^{\rm{out}}$ of 0.2. For visual clarity, regions with $\Delta P_c^{\rm{out}}>0$ are marked with a '+', and regions where $\Delta P_c^{\rm{out}}<0$ are marked with a '-'.}
\label{fig:pout}
\end{figure*}

We now visualise how the outgoing critical pressure depends on the four parameters $\theta_1$, $\theta_2$, $\alpha$, and $R_2'$. To reduce the dimensionality of the representation, in Fig. \ref{fig:pout} we show a matrix of contour plots at fixed $\theta_1$ and $\theta_2$, both of which may only take the values $10^\circ$, $50^\circ$, and $100^\circ$. We choose these values to capture the range of contact angles exhibited by commonly used liquids and substrates, see for example \cite{Lee2008}. 


At each $\theta_1, \theta_2$, Fig. \ref{fig:pout} illustrates the sets of critical pressure states presented in Table \ref{table:outtypes}. We now discuss the competition between the states within each set, for the global critical pressure. 

When $\theta_1=10^\circ$, $\alpha<\theta_2$, B, B', I, or I' are possible, however the I state is not observed within the range of $R_2'$ plotted. The region of existence of B' is shown to not depend on $\theta_2$. This is because the B' meniscus is pinned to the bottom of the well, and so never experiences the chemical gradient. Furthermore, if B' co-exists with I', the I' state must necessarily have a lower critical pressure than B'. This is because, compared to B', the I' meniscus has a wider contact line radius and smaller peak height, leading to a greater radius of curvature and so a smaller critical pressure. Thus, the I' critical pressure (which does depend on $\theta_2$) never out-competes the B' critical pressure, leaving the B' region of existence unaltered by $\theta_2$. I' is however able to out-compete the B state, as exhibited by the B region receding to larger $R_2'$ values as $\theta_2$ is increased from $10^\circ$ to $50^\circ$.

When $\theta_1=50^\circ$, two additional features are observed. The first is that the B and T states only coexist and compete when $\theta_1>\theta_2$ (and $\theta_1,\theta_2<\alpha$ as described in Table \ref{table:outtypes}, meaning the menisci at the top and bottom of the pore are concave). This condition must be satisfied, otherwise the wider aperture at the top of the pore will always produce a meniscus of larger negative critical radius, and so a greater critical pressure, than when pinned to the bottom of the pore. The second additional feature is that for $\theta_2=100^\circ$, we now observe the I state to exist over a small region at large $R_2'$. For I to occur, the critical meniscus must be produced sufficiently low in the well for the peak to not contact the upper substrate. This requires a delicate balance between the chemical gradient (favouring the critical state at the top of the pore) and the physical gradient (favouring the critical state at the bottom of the pore), which overall produces a narrow existence range of I. 

When $\theta_1=100^\circ$, only the B and B' states are able to occur, as outlined in Table \ref{table:outtypes}. Because the contact line at the critical pressure is always pinned to the bottom of the well, $\theta_2$ has no impact on the critical pressure. Thus, all three contour plots for $\theta_1=100^\circ$ are identical. Since $\theta_1>90^\circ$, we also observe here the incidence of the maximum possible critical pressure, $\Delta P^{\rm{out}}=1$, shown bounded by the thick contour. This is shown in Eq. \eqref{eqn:P_B} to be as a result of the B-state critical pressure occurring when the contact line is pinned to the bottom of the well, with a contact angle of $90^\circ$ with respect to the horizontal axis.

\subsection*{Incoming critical pressures}
\label{sec:in}
\subsubsection*{Setup}

\begin{figure}[!h]
\includegraphics[width=0.5\textwidth]{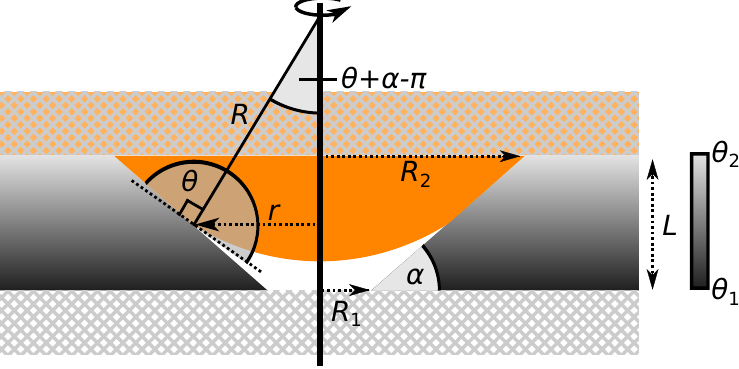}
\caption{2D illustration of the axisymmetric pore construction and incoming meniscus profile, the axis of symmetry shown as the thick vertical line. Liquid is shown in orange, with air shown in white.}
\label{fig:insetup}
\end{figure}

We model the occurrence of liquid entering the pore from above in Fig. \ref{fig:insetup}. We utilise the same setup as shown in Fig. \ref{fig:outsetup}, with the exception that the liquid (orange) now enters from the top absorbent substrate, and the bottom substrate is dry. The same linear physical and chemical gradients are employed as before. Again, as we focus on modelling the maximum pressure maintainable before fluid enters the pore, the contact angles used in the analysis are the advancing contact angles on a surface where hysteresis is present.

\begin{figure}[!ht]
\includegraphics[width=0.5\textwidth]{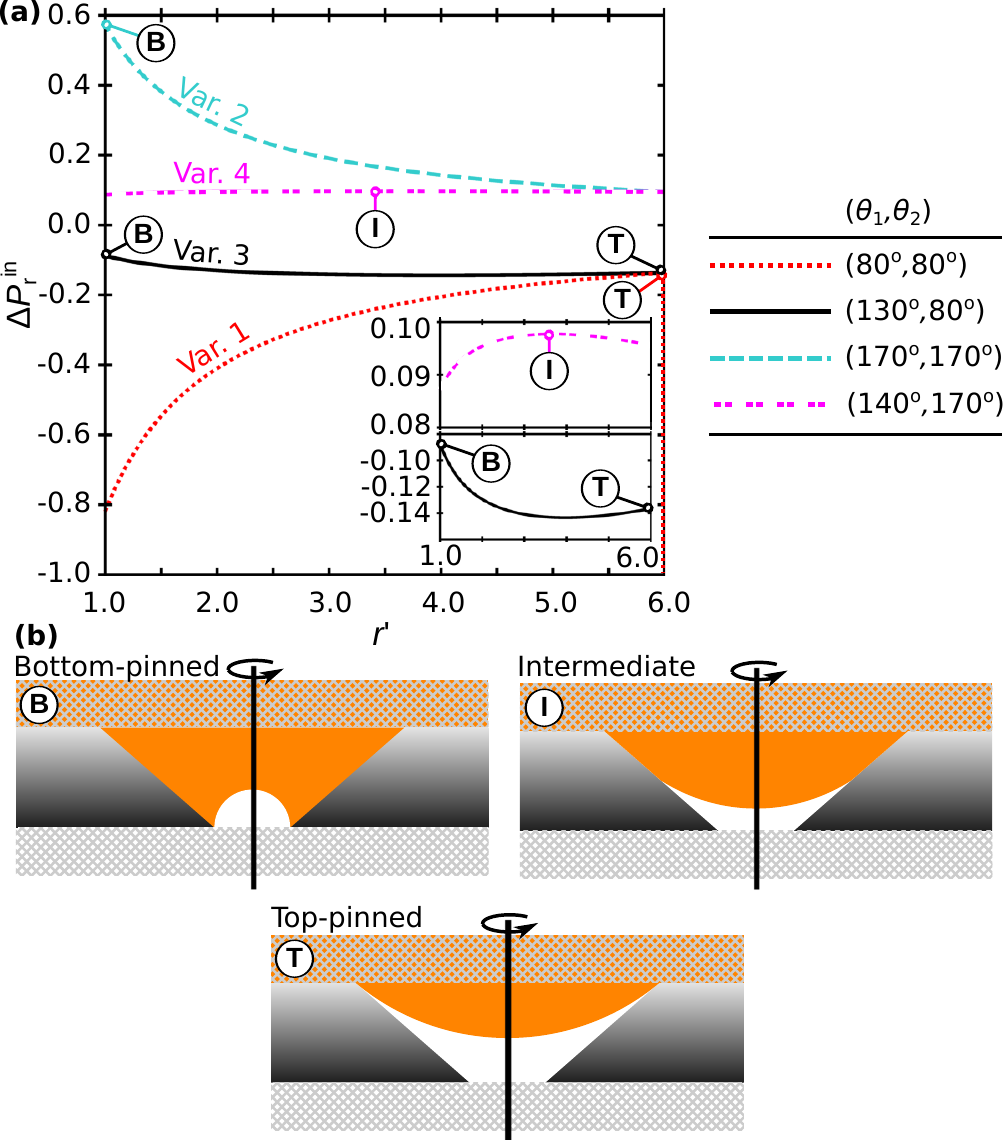}
\caption{(a) Example plots of each of the four incoming reduced pressure variations with $r'$, with insets magnifying the local maxima/minima. The local maxima of each variation is associated with one of three critical meniscus types, illustrated in (b).}
\label{fig:intypes}
\end{figure}

The incoming Laplace pressure $\Delta P_r^{\rm{in}}$ can be derived as
\begin{equation}
\Delta P^{\rm{in}}_r = -\frac{1}{r'} \sin \left[ \theta_1 + (\theta_2-\theta_1)\frac{r'-1}{R_2'-1} + \alpha \right]. \label{eqn:DELTA_P_IN}
\end{equation}
It can be seen that Eq. \eqref{eqn:DELTA_P_IN} can be obtained from $\Delta P^{\rm{out}}_r$ in Eq. \eqref{eqn:DELTA_P_OUT} by exchanging the fluid phases, such that $\theta(r') \rightarrow \pi - \theta(r')$. This simple transformation however will be shown to give rise to remarkably different incoming and outgoing critical pressures. Again, the competition between physical and chemical gradients gives rise to four different variations in $\Delta P_r^{\rm{in}}$ with $r'$, where we have reserved the study of the interaction between the meniscus apex and lower absorbent substrate to the section entitled, "Influence of top substrate: T' and I'". The characteristic examples of each shown in Fig. \ref{fig:intypes}(a) illustrate the symmetry between $\Delta P^{\rm{in}}_r$ and $\Delta P^{\rm{out}}_r$, as it is observed that by making the fluid exchange,  Fig. \ref{fig:outtypes}(a) is reflected in the r' axis to yield Fig. \ref{fig:intypes}(a). 

\subsubsection*{Variation 1 and the top-pinned (T) critical meniscus}
The first variation is shown as the dotted red line in Fig. \ref{fig:intypes}(a). Here, for $r' < R_2'$, $\Delta P^{\rm{in}}_r$ decreases monotonically as $r'$ decreases. When $r' = R_2'$ however, a range of critical pressures is possible as the contact line is pinned to the top of the pore. This is shown as the vertical dotted red line at $r' = R_2'$. The critical pressure here occurs in the top-pinned state, labelled the T-state in Fig. \ref{fig:intypes}(b), where,
\begin{equation}
\Delta P^{\rm{in}}_c(\rm{T})=-\frac{1}{\mathit{R}_2'}\sin\left( \theta_2+\alpha \right). \label{eqn:INP_T}
\end{equation}

\subsubsection*{Variation 2 and the bottom-pinned (B) critical meniscus} 
The second variation is shown as the dashed cyan line in Fig. \ref{fig:intypes}(a). This shows a monotonic increase of $\Delta P^{\rm{in}}_r$ as $r'$ is decreased. The critical pressure therefore occurs at the point when the contact line reaches the bottom of the system at $r'=1$, $\theta = \theta_1$, labelled the B-state in Fig. \ref{fig:intypes}(b). In this case,
\begin{equation}
\Delta P^{\rm{in}}_c(\rm{B})= -\sin\left( \theta_1 + \alpha \right). \label{eqn:INP_B}
\end{equation}

\subsubsection*{Variation 3 and the T and B critical menisci}
The third variation is shown as the solid black line in Fig. \ref{fig:intypes}(a). Here, a local minimum exists at intermediate values of  $r'$. The lower inset panel highlights the local minimum in a vertical magnification. In this pressure variation, both the T-state and the B-state become local maximisers of $\Delta P^{\rm{in}}_r$. Which state globally maximises $\Delta P^{\rm{in}}_r$ is found by comparing Eq. \eqref{eqn:INP_T} and Eq. \eqref{eqn:INP_B}. We perform this comparison for selected examples of $\theta_1$ and $\theta_2$ in the Incoming critical pressures visualisation section.

\subsubsection*{Variation 4 and the intermediate (I) critical meniscus}
The fourth variation is shown as the double-dashed magenta line in Fig. \ref{fig:intypes}(a). Here, a local maximum is obsetved at intermediate values of  $r'$, labelled the I-state in Fig. \ref{fig:intypes}(b). The upper inset panel in Fig. \ref{fig:intypes}(a) highlights the local maximum in a vertical magnification. This local maximum can be found by finding stationary points of $\Delta P^{\rm{in}}_r$ in Eq. \eqref{eqn:DELTA_P_IN} which maximise $\Delta P^{\rm{in}}_r$ in the interval $r_c' \in [1, R_2']$. This is achieved through solving,
\begin{align}
&\frac{1}{r_c'}\left(\frac{\theta_2-\theta_1}{R_2'-1}\right)\cos\left(\theta_1+(\theta_2-\theta_1)\frac{r_c'-1}{R_2'-1}+\alpha\right) \nonumber \\ 
 -&\frac{1}{r_c'^2}\sin\left(\theta_1+(\theta_2-\theta_1)\frac{r_c'-1}{R_2'-1}+\alpha\right) = 0. \label{eqn:INP_I}
\end{align}
In general, this is again not analytically solvable and instead must be solved numerically.

\subsubsection*{Influence of top substrate: T' and I'}
When the liquid meniscus is convex, the centre of meniscus may contact the bottom of the pore before the T-state or I-state critical pressure is reached. 

\begin{figure}[!ht]
\includegraphics[width=0.5\textwidth]{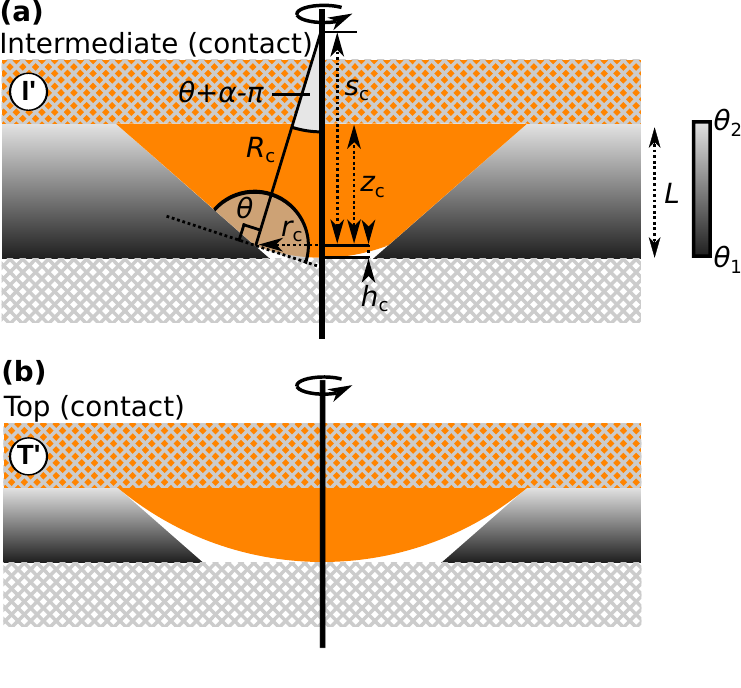}
\caption{(a) Construction used to calculate the critical pressure of the I' state. (b) Illustration of the B' state, with the point of failure highlighted by a red circle.}
\label{fig:intypesprime}
\end{figure}

To find the critical pressure of the I-contacting state shown in Fig. \ref{fig:intypesprime}(a) (denoted I'), we begin by finding the total sag depth of the liquid meniscus. This is constructed as the sum of the depth of the contact line $z_c$ below the pore top, and the depth of the spherical cap below this $h_c$. Using $h_c = R_c - s_c$,  we derive,
\begin{equation}
h_c=-r_c\frac{1+\cos(\theta(r_c)+\alpha)}{\sin(\theta(r_c)+\alpha)}.
\end{equation}
The spherical cap will touch the lower substrate if $z_c+h_c=L$. In reduced units, we therefore solve,
\begin{equation}
\left(1-r_c'\right)\tan\alpha - r_c'\frac{1+\cos(\theta(r_c')+\alpha)}{\sin(\theta(r_c')+\alpha)}=0.
\end{equation}
In general this does not have analytic solutions and must be solved numerically. Once $r_c'$ is found in this way, we substitute $r'$ for $r_c'$ in Eq. \eqref{eqn:DELTA_P_IN} to obtain the critical pressure $\Delta P^{\rm{in}}_c(\rm{I'})$ caused by the cap contacting the lower substrate, while the contact line radius takes an intermediate value between $R_1$ and $R_2$. 

If instead the contact line remains pinned to the top of the pore at the point of meniscus contact, as illustrated in Fig. \ref{fig:intypesprime}(b), the top-contacting critical meniscus arises (denoted T'), where the incoming critical pressure is expressed as,
\begin{equation}
\Delta P^{\rm{in}}_c(\rm{T'})= \frac{2}{\mathit{L}'+\frac{\mathit{R}_2'^2}{\mathit{L}'}}. \label{eqn:INP_TPRIME}
\end{equation}

\subsubsection*{Critical morphology existence ranges}
Overall, five different critical interface morphologies may occur: B, T, T', I, and I', in which the associated critical pressures feature different dependencies on $\theta_1$, $\theta_2$, $\alpha$, and $R_2'$. Despite this, the critical pressure types able to occur within a system can be determined based on whether the liquid meniscus is convex ($\theta + \alpha > \pi$) or concave ($\theta + \alpha < \pi$) and the top and bottom of the system. This is presented in Table \ref{table:intypes}.
\begin{table}[!h]
\small
\caption{The critical incoming meniscus types able to occur for a convex meniscus, $\theta+\alpha>\pi$, or concave meniscus $\theta+\alpha<\pi$.}
\label{table:intypes}
	\begin{tabular*}{0.48\textwidth}{@{\extracolsep{\fill}}lll}
	\hline
	& $\theta_1+\alpha>\pi$ & $\theta_1+\alpha<\pi$ \\
	\hline
	$\theta_2+\alpha>\pi$ & T, T', I, I' & T, T' \\
	$\theta_2+\alpha<\pi$ & I, I' & T, B \\
	\hline
	\end{tabular*}
\end{table}

For ($\theta_1+\alpha<\pi$, $\theta_2+\alpha>\pi$), the critical meniscus must occur when contact line is pinned to the top of the system as T or T'. For ($\theta_1+\alpha>\pi$, $\theta_2+\alpha<\pi$) however, the Laplace pressure is negative when the contact line is at the top of the pore, and positive at the bottom, so the critical meniscus must occur in some intermediate state: I or I'. For ($\theta_1+\alpha>\pi$, $\theta_2+\alpha>\pi$), the meniscus is convex for all $r$, meaning the T, T', I, or I' states could occur. For ($\theta_1+\alpha<\pi$, $\theta_2+\alpha<\pi$), the meniscus is concave for all $r$, so that the critical pressure must either occur at the top of the system, as T, or bottom, as B.

\subsubsection*{Incoming critical pressure visualisation}
\label{sec:in_vis}

\begin{figure*}[!ht]
\centering
\includegraphics[width=\textwidth]{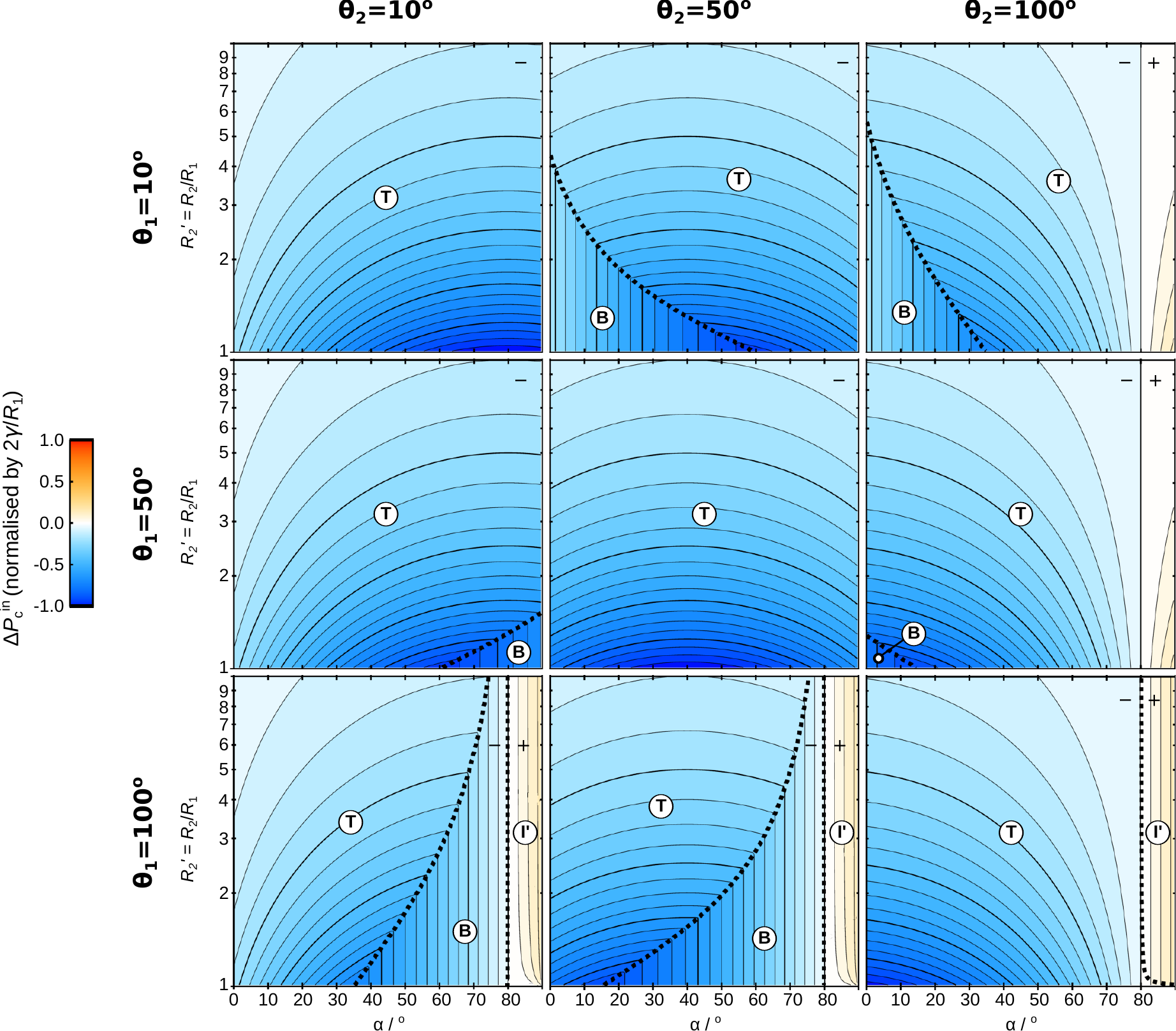}
\caption{Matrix of $R_2'$ - $\alpha$ contour plots of the incoming critical pressure for a selection of $\theta_1$ and $\theta_2$. The incoming meniscus types are labelled with white circles. The boundaries between these critical types are shown as dotted black lines. Contours are shown at intervals in $\Delta P_c^{\rm{in}}$ of 0.05. For visual clarity, regions with $\Delta P_c^{\rm{in}}>0$ are marked with a '+', and regions where $\Delta P_c^{\rm{in}}<0$ are marked with a '-'.}
\label{fig:pin}
\end{figure*}

We now visualise how the incoming critical pressure depends on the four parameters $\theta_1$, $\theta_2$, $\alpha$, and $R_2'$. To reduce the dimensionality of the representation, in Fig. \ref{fig:pin} we again show a matrix of contour plots at fixed $\theta_1$ and $\theta_2$, both of which may only take the values $10^\circ$, $50^\circ$, and $100^\circ$.

The incoming critical pressure contour plots in Fig. \ref{fig:pin} show markedly different behaviour to the outgoing critical pressure plots in Fig. \ref{fig:pout}. This is because under the range of $\theta_1, \theta_2$ tested, except at large $\alpha$, $\theta_1 + \alpha < \pi$ and $\theta_2 +\alpha <\pi$ meaning the liquid meniscus is concave. This means that the pore exerts a pulling force on the liquid in the top substrate, so that to prevent liquid filling the pore, a negative pressure must be applied. 

\begin{figure*}[!ht]
\centering
\includegraphics[width=\textwidth]{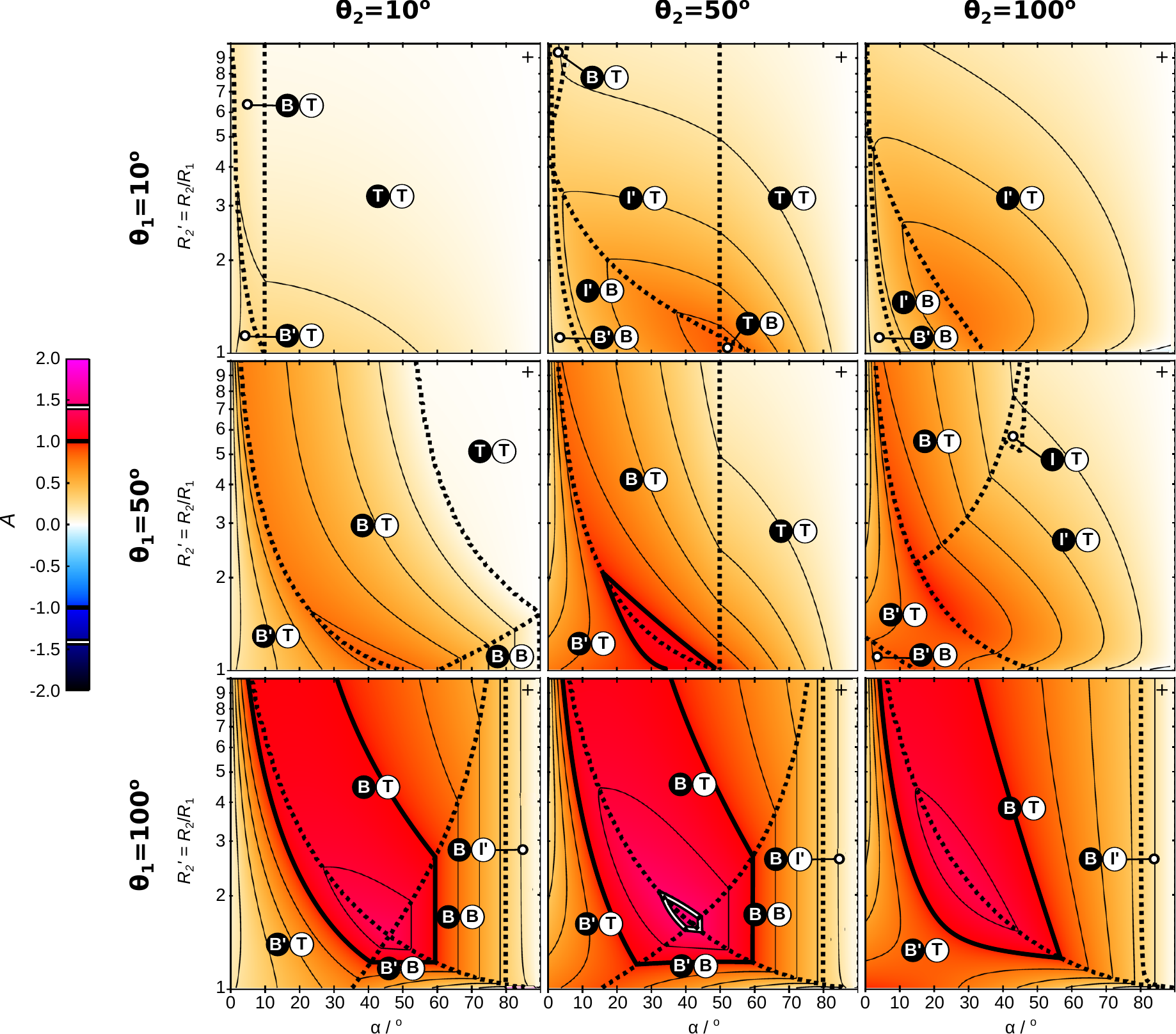}
\caption{Matrix of $R_2'$ - $\alpha$ contour plots of the critical pressure asymmetries for a selection of $\theta_1$ and $\theta_2$. The outgoing meniscus incoming meniscus pair types are labelled with black and white circles. The boundaries between these critical types are shown as dotted black lines. Contours are shown at intervals in $A$ of 0.2. Two significant $A$ contours are also highlighted, $A=1$ (thick black line), and $A=\sqrt{2}$ (white-centred black line). }
\label{fig:delta}
\end{figure*}

Under such conditions, when $\theta_1 = \theta_2$, the T-type critical meniscus always emerges. This is because the meniscus has a wider contact radius at the top of the pore than the bottom, resulting in the less-negative critical pressure at the top of the pore. When $\theta_1 > \theta_2$, the B-state may out-compete the T-state for largest critical pressure at large $\alpha$ and $R_2'$. When this happens, the high contact angle at the bottom of the pore negates the small contact line radius, to create a less-negative critical pressure than the T-state. When $\theta_1 < \theta_2$, this time the B-state may out-compete the T-state for largest critical pressure at small $\alpha$ and $R_2'$. This is because at the bottom of the pore, the low contact angle creates a liquid-vapour interface with a near-spherical shape. The associated large radius of curvature produces a smaller negative critical pressure than the meniscus at the top of the well.

When $\alpha$ is sufficiently large to enable $\theta_1 + \alpha > \pi$, the I' state is observed. This is because a convex meniscus is enabled close to the bottom of the well, resulting in a positive critical pressure.

\subsection*{Critical pressure asymmetry}
\label{sec:asym}

We now define the critical pressure asymmetry, \textit{A}, of a pore: the difference between the outgoing and incoming critical pressures,
\begin{equation}
A=\Delta P^{\rm{out}}_r-\Delta P^{\rm{in}}_r. \label{eqn:deltap}
\end{equation}
A matrix of contour plots shown in Fig. \ref{fig:delta} illustrates the rich and complex dependence of \textit{A} on $\theta_1$, $\theta_2$, $\alpha$, and $R_2'$. We identify three important values of $A$ to consider, which are deduced in Supporting Information (Fig. S1). $A=\pm1$, shown as the thick contour, is the maximum possible asymmetry for a doubly-closed cylindrical pore when $R_2 \rightarrow \infty$. $A=\pm\sqrt{2}$ shown as the doubly-thick contour, is the maximum possible asymmetry for a semi-open cylindrical pore. $A=2$ is the maximum possible asymmetry for any pore. Recently a semi-open pore has been developed which approaches this maximum \cite{Agonafer2018}. We now discuss the features of the critical pressure asymmetries.

\begin{figure*}[!t]
\centering
\includegraphics[width=\textwidth]{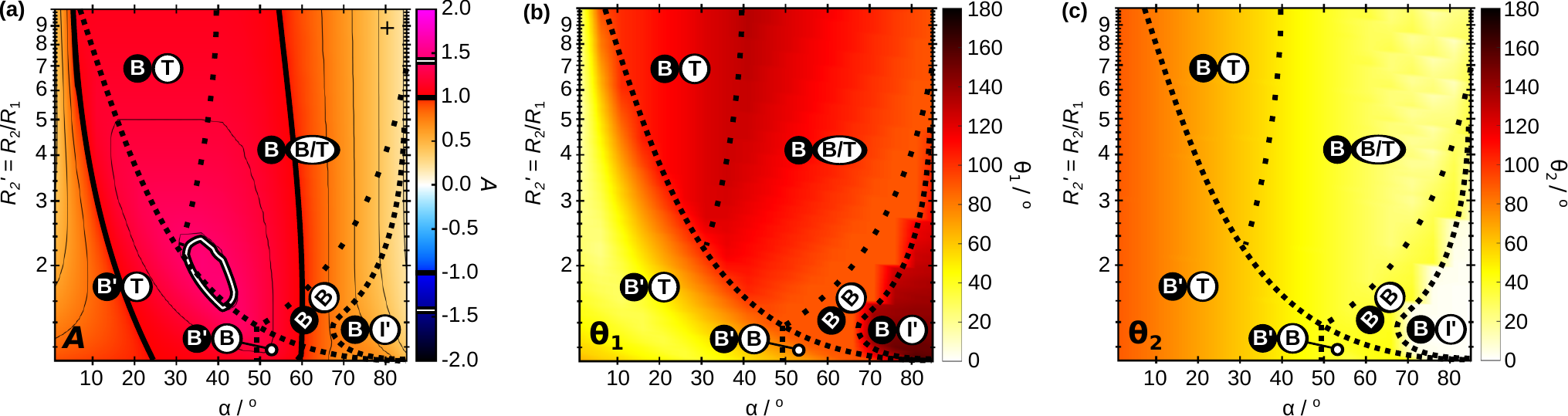}
\caption{(a) Maximum asymmetry possible at each ($\alpha$, $R_2'$) coordinate. Two significant $A$ limits are highlighted in the contour plots, $A=1$ (thick black line), and $A=\sqrt{2}$ (white-centred black line). The outgoing meniscus incoming meniscus pair types are labelled with black and white circles, with dense dotted black lines showing the boundaries between these types. The diffuse black dotted line illustrates the approximate region where the incoming B and T states have equal pressure. (b) $\theta_1$ required for maximum asymmetry. (c) $\theta_2$ required for maximum asymmetry.}
\label{fig:opt}
\end{figure*}

It is initially observed that for the range of parameters shown in Fig. \ref{fig:delta}, $A$ is never negative. This can be proved to be true in general for all $\alpha < \pi/2$, shown in the Supporting Information (Fig. S2). Thus, for $\alpha < \pi/2$, a conical pore will always preferentially intake liquid than expel it.

For $\theta_1=10^\circ$, the critical pressure asymmetry remains small. When $\theta_1<\alpha$, $A$ is small because both the incoming and outgoing critical pressures are negative (with the magnitude of the outgoing being smaller than the incoming). When $\theta_1>\alpha$, $A$ is also small as although the outgoing critical pressure is positive, it is never large. This is because $\Delta P^{\rm{out}}_c(\rm{B})$ is small due to the low contact angle, and $\Delta P^{\rm{out}}_c(\rm{I'})$ is small due to the large contact radius. Overall, the asymmetry is dominated by the negative contribution from the incoming critical pressure, rather than the outgoing critical pressure. Because of the dominance of the incoming critical pressure, the maximal $A$ occurs on the boundary between the incoming B and T critical meniscus types.

For $\theta_1=50^\circ$, a similar picture emerges. However, now for $\theta_1>\alpha$, the B- or B'-type outgoing critical pressure is able to be large and positive. In this region, we therefore begin to see larger $A$ as the contribution of the outgoing critical pressure becomes more significant. The competition also becomes apparent between intermediate positive outgoing critical pressures (at small $\alpha$, large $R_2'$), and large negative incoming critical pressures (at intermediate $\alpha$, small $R_2'$). The maximum asymmetries occur as an optimal compromise between these extremes, at intermediate $\alpha$ and $R_2'$. In contrast to $\theta_1=10^\circ$, the maximum asymmetries now occur along an outgoing boundary.

For $\theta_1=100^\circ$, very large asymmetries are observed, exceeding $A=1$ in all panels examined, and exceeding $A=\sqrt{2}$ when $\theta_2=50^\circ$. Here, the large $\theta_1$ enables large $\Delta P^{\rm{out}}_c(\rm{B/B'})$. Thus, the competition between large outgoing and incoming critical pressures observed for $\theta_1=50^\circ$ becomes here more extreme. Interestingly, now that both the outgoing and incoming critical pressures are able to contribute equally to $A$, the maximum asymmetries are observed at points where an outgoing-type boundary and incoming-type boundary cross. This is most apparent when $\theta_1=100^\circ$, $\theta_2=50^\circ$, where at the junction between the outgoing B/B' boundary and incoming T/B boundary, $A>\sqrt{2}$.

\subsection*{Optimum asymmetry}
\label{sec:opt}

\begin{figure*}[!ht]
\centering
\includegraphics[width=0.72\textwidth]{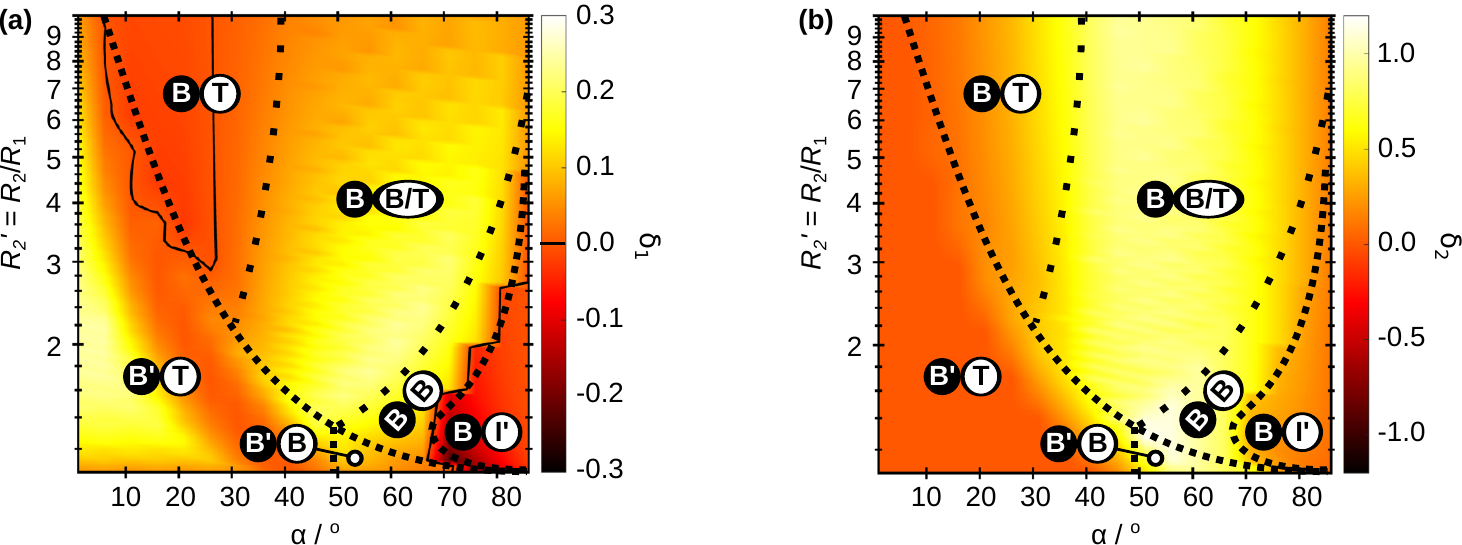}
\caption{Comparison of the asymmetry of the optimal chemically patterned conical pore, with the sum of asymmetries from the same chemical patterning applied to a cylindrical pore, and a chemically homogeneous conical pore. The comparison is made for the chemically homogeneous conical pore with $\theta = \theta_1^{\rm{opt}}$ in (a), and $\theta = \theta_2^{\rm{opt}}$ in (b). Solid black contours mark the $\delta=0$ level. Note the larger scale bar range in (b).}
\label{fig:sum_parts}
\end{figure*}

We observe in Fig. \ref{fig:delta} that the ($\alpha$, $R_2'$) coordinate that maximises the asymmetry depends on both $\theta_1$ and $\theta_2$. We now investigate the maximum possible asymmetry for a specified $\alpha$ and $R_2'$, by varying $\theta_1$ and $\theta_2$. This is achieved by evaluating the asymmetry at each ($R_2'$,$\alpha$)  coordinate, when $\theta_1$ and $\theta_2$ are iteratively incremented in 2$^\circ$ steps from 0$^\circ$ to 180$^\circ$. The overall optimum asymmetry is shown in Fig. \ref{fig:opt}(a), with the associated optimal $\theta_1$ and $\theta_2$ shown in Fig. \ref{fig:opt}(b) and Fig. \ref{fig:opt}(c) respectively.

Overall, we can conclude that the maximum possible asymmetry, $A=$ 1.46,  occurs at:  $\alpha=$ 41$^\circ$ $\pm$ 5$^\circ$, $R_2'=$ 1.70 $\pm$ 0.05, $\theta_1=$ 104$^\circ$ $\pm$ 2$^\circ$ and $\theta_2=$ 49$^\circ$ $\pm$ 2$^\circ$. Uncertainties reported indicate the resolution with which the quantities were determined. Across all $\alpha$ and $R_2'$, a key trend we observe is that a high contrast between $\theta_1$ and $\theta_2$ is required in order to produce maximum asymmetries (a homogeneous contact angle will not in general maximise $A$).  We also observe that  $\theta_1$ and $\theta_2$ vary non-monotonically with both $\alpha$ and $R_2'$, due to changes in the critical meniscus type.

We now examine the critical meniscus types observed to achieve maximum asymmetry. For the outgoing critical pressure, the observed strategy for maximising $A$ is to ensure the contact line remains pinned to the bottom of the pore in the B or B' 
state, thereby ensuring $\Delta P_c^{\rm{out}}$ remains large and positive. 

The incoming critical meniscus type is however more variable, particularly in the region where the outgoing critical meniscus is the B-state in the upper right-hand sides of the plots in Fig. \ref{fig:opt}. Here, the incoming meniscus may be in the I', T or B states. In the region where I' is dominant, the incoming critical pressure is positive, making this region unique across all $\alpha$ and $R_2'$. To maximise the asymmetry in this region, the most effective strategy is to maximise $\Delta P_c^{\rm{out}}$ using $\theta_1 \approx \alpha + \pi/2$, at the expense of enabling a positive incoming critical pressure. However, the large value of $\alpha$ ensures $\Delta P_c^{\rm{in}}$ is never too large, and is minimised further by setting $\theta_2 = 0^\circ$.

When the I' state does not occur however, the B and T incoming states compete for the largest negative $\Delta P_c^{\rm{in}}$. As is observed in Fig. \ref{fig:pin}, the largest negative critical pressures occur on the boundary between B and T (where this boundary exists). Thus, over the extended region outlined with diffuse dotted lines in Fig. \ref{fig:opt}, the incoming critical pressure is equally described by both the B and T states.

Finally, we examine the cooperativity  of the chemical and physical gradients in producing the critical pressure asymmetry. We do this by comparing the optimal critical pressure asymmetry to 'the sum of its parts': the cooperativity $\delta$ is defined as
\begin{equation}
\delta = A - \left( A_{\rm{het}}^{\rm{cyl}} + A_{\rm{hom}}^{\rm{con}}  \right).
\end{equation}
At each ($R_2'$, $\alpha$) coordinate, $A_{\rm{het}}^{\rm{cyl}}$ is evaluated as the critical pressure asymmetry of a cylindrical pore, with a chemical gradient the same as the optimal chemical gradient shown in Figs. \ref{fig:opt}(b) and (c). For a fair comparison, we also ensure the pore depth $L$ is the same for the cylindrical and conical pores at each  ($R_2'$, $\alpha$) coordinate. $A_{\rm{hom}}^{\rm{con}}$ is the critical pressure asymmetry for a chemically homogeneous conical pore. A number of choices exist in deciding which homogeneous contact angle most fairly compares to the optimal chemical gradient. We show in Fig. \ref{fig:sum_parts} the two limiting cases, when the homogeneous contact angle is: (i) the optimal $\theta_1$ at each ($R_2'$, $\alpha$) coordinate; (ii) the optimal $\theta_2$ at each ($R_2'$, $\alpha$) coordinate. To distinguish these two limiting cooperativities, we label these $\delta_1$ in Fig. \ref{fig:sum_parts}(a), and $\delta_2$ in Fig. \ref{fig:sum_parts}(b) respectively.

In Fig. \ref{fig:sum_parts}(a), over the majority of the ($R_2'$, $\alpha$) plane, the cooperativity $\delta_1 > 0$, meaning that the optimal asymmetry (arising from both physical and chemical gradients) is greater than the sum of asymmetries arising from the physical gradient and chemical gradient separately. Thus, the physical and chemical gradients act together to produce the high optimal asymmetries. The exceptions to this, when $\delta_1 < 0$ within the solid black contours, arise when the optimal asymmetry is almost wholly achieved through the conical shape and not the chemical patterning. As  $A_{\rm{het}}^{\rm{cyl}} >0$, in these cases, the optimal asymmetry is less than the sum of its parts.

In Fig. \ref{fig:sum_parts}(b), we observe an extended region at intermediate values of $\alpha$ for which the cooperativity $\delta_2 > 1$. This very large, positive cooperaitvity is caused in this region by the occurrence of the I-type outgoing critical meniscus for the homogeneous conical pore. The I-type outgoing critical pressure is smaller than the B- or B'-types, hence $A_{\rm{hom}}^{\rm{con}}$ is small, leading to the large $\delta_2$ observed. The impact of B- or B'-type outgoing critical pressures instead of I can be seen in Fig. \ref{fig:sum_parts}(a). Here, the outgoing type is always B or B', leading to a larger $A_{\rm{hom}}^{\rm{con}}$, and hence a smaller $\delta_1$.

\section*{Summary and conclusions}
Here we have calculated the maximum Laplace pressures (the critical pressures) required for fluid to both enter and leave a conical, chemically-patterned pore, sandwiched between two absorbent substrates. Across the range of pore designs considered, we found the Laplace pressure to depend on the contact line radius in four different manners; of which two of these arose from a competition between the physical and chemical gradients. This interaction between the two gradients produced three different critical menisci, where the contact line was: pinned to top of the pore, pinned to the bottom of the pore, or located in between. The presence of the top and bottom substrates produced an additional two critical menisci, due to premature contact of the liquid-vapour interface with the substrates.

We then analysed the critical pressure asymmetry, the difference between incoming and outgoing critical pressures, as a measure of the efficacy of the fluid diode across a range of pore geometries. For the pores considered with an opening angle $\alpha < 90^{\circ}$, the outgoing pressure was always shown to be larger than the incoming pressure. Furthermore, the maximum asymmetry did not in general occur due to the dominance of either the incoming or outgoing critical pressure individually, but as a compromise between the two. 

Finally, we optimised the chemical pattering to produce maximal critical pressure asymmetries across the range of pore geometries, showing that a large chemical gradient is required to produce large asymmetries. Across the majority of pore opening angles $\alpha$ and maximum radii $R_2$, we showed that the optimum asymmetry for the pore with both physical and chemical gradients was greater than the sum of asymmetries of pores with physical gradients and chemical gradients separately. The physical and chemical gradients therefore act together cooperatively to achieve the largest critical pressure asymmetries.

\section*{Acknowledgements}
The authors would like to thank Procter \& Gamble for funding.

\section*{Supporting information}
Derivation of the critical pressure asymmetry limits and proof that for $\alpha<\pi/2$, the asymmetry $A>0$.

\normalem
\bibliography{Bibliography}

\begin{thebibliography}{10}

\bibitem{Li2019}
J.~Li, J.~Li, J.~Sun, S.~Feng, and Z.~Wang, ``{Biological and Engineered
  Topological Droplet Rectifiers},'' {\em Advanced Materials}, vol.~31, no.~14,
  p.~1806501, 2019.

\bibitem{Ju2012}
J.~Ju, H.~Bai, Y.~Zheng, T.~Zhao, R.~Fang, and L.~Jiang, ``{A multi-structural
  and multi-functional integrated fog collection system in cactus.},'' {\em
  Nature communications}, vol.~3, p.~1247, 2012.

\bibitem{Bai2010}
H.~Bai, X.~Tian, Y.~Zheng, J.~Ju, Y.~Zhao, and L.~Jiang, ``{Direction
  Controlled Driving of Tiny Water Drops on Bioinspired Artificial Spider
  Silks},'' {\em Advanced Materials}, vol.~22, no.~48, pp.~5521--5525, 2010.

\bibitem{Kusumaatmaja2009}
H.~Kusumaatmaja and J.~M. Yeomans, ``{Anisotropic hysteresis on ratcheted
  superhydrophobic surfaces},'' {\em Soft Matter}, vol.~5, no.~14, p.~2704,
  2009.

\bibitem{Renvoise2009}
P.~Renvois{\'{e}}, J.~W.~M. Bush, M.~Prakash, and D.~Qu{\'{e}}r{\'{e}}, ``{Drop
  propulsion in tapered tubes},'' {\em EPL (Europhysics Letters)}, vol.~86,
  no.~6, p.~64003, 2009.

\bibitem{Lorenceau2004}
L.~Lorenceau and D.~Qu{\'{e}}r{\'{e}}, ``{Drops on a conical wire},'' {\em
  Journal of Fluid Mechanics}, vol.~510, pp.~29--45, 2004.

\bibitem{Brochard1989}
F.~Brochard, ``{Motions of droplets on solid surfaces induced by chemical or
  thermal gradients},'' {\em Langmuir}, vol.~5, no.~2, pp.~432--438, 1989.

\bibitem{Gennes2010}
P.-G. de~Gennes, F.~Brochard-Wyart, and D.~Qu{\'{e}}r{\'{e}}, {\em Capillarity
  and Wetting Phenomena: Drops, Bubbles, Pearls, Waves}.
\newblock New York, USA: Springer Science + Business Media, Inc., 2010.

\bibitem{Mates2014}
J.~E. Mates, T.~M. Schutzius, J.~Qin, D.~E. Waldroup, and C.~M. Megaridis,
  ``{The fluid diode: Tunable unidirectional flow through porous substrates},''
  {\em ACS Applied Materials and Interfaces}, vol.~6, no.~15, pp.~12837--12843,
  2014.

\bibitem{Zhang2019}
S.~Zhang, J.~Huang, Z.~Chen, S.~Yang, and Y.~Lai, ``{Liquid mobility on
  superwettable surfaces for applications in energy and the environment},''
  {\em Journal of Materials Chemistry A}, vol.~7, no.~1, pp.~38--63, 2019.

\bibitem{Zhao2017}
Y.~Zhao, H.~Wang, H.~Zhou, and T.~Lin, ``{Directional Fluid Transport in Thin
  Porous Materials and its Functional Applications},'' {\em Small}, vol.~13,
  no.~4, p.~1601070, 2017.

\bibitem{Brown2016}
P.~S. Brown and B.~Bhushan, ``{Bioinspired materials for water supply and
  management: water collection, water purification and separation of water from
  oil},'' {\em Philosophical Transactions of the Royal Society A: Mathematical,
  Physical and Engineering Sciences}, vol.~374, no.~2073, p.~20160135, 2016.

\bibitem{Li2017}
J.~Li, X.~Zhou, J.~Li, L.~Che, J.~Yao, G.~McHale, M.~K. Chaudhury, and Z.~Wang,
  ``{Topological liquid diode},'' {\em Science Advances}, vol.~3, no.~10,
  p.~eaao3530, 2017.

\bibitem{Shou2018}
D.~Shou and J.~Fan, ``{An All Hydrophilic Fluid Diode for Unidirectional Flow
  in Porous Systems},'' {\em Advanced Functional Materials}, vol.~28, no.~36,
  p.~1800269, 2018.

\bibitem{Zimmermann2008}
M.~Zimmermann, P.~Hunziker, and E.~Delamarche, ``{Valves for autonomous
  capillary systems},'' {\em Microfluidics and Nanofluidics}, vol.~5, no.~3,
  pp.~395--402, 2008.

\bibitem{Diersch2010}
H.-J.~G. Diersch, V.~Clausnitzer, V.~Myrnyy, R.~Rosati, M.~Schmidt, H.~Beruda,
  B.~J. Ehrnsperger, and R.~Virgilio, ``{Modeling Unsaturated Flow in Absorbent
  Swelling Porous Media: Part 1. Theory},'' {\em Transport in Porous Media},
  vol.~83, no.~3, pp.~437--464, 2010.

\bibitem{Miao2018}
D.~Miao, Z.~Huang, X.~Wang, J.~Yu, and B.~Ding, ``{Continuous, Spontaneous, and
  Directional Water Transport in the Trilayered Fibrous Membranes for
  Functional Moisture Wicking Textiles},'' {\em Small}, vol.~14, no.~32,
  pp.~1--10, 2018.

\bibitem{Shi2018}
L.~Shi, X.~Liu, W.~Wang, L.~Jiang, and S.~Wang, ``{A Self-Pumping Dressing for
  Draining Excessive Biofluid around Wounds},'' {\em Advanced Materials},
  vol.~1804187, p.~1804187, 2018.

\bibitem{Cho2007}
H.~Cho, H.-Y. Kim, J.~Y. Kang, and T.~S. Kim, ``{How the capillary burst
  microvalve works},'' {\em Journal of Colloid and Interface Science},
  vol.~306, pp.~379--385, 2007.

\bibitem{Chen2008}
J.~M. Chen, C.-Y. Chen, and C.-H. Liu, ``{Pressure Barrier in an Axisymmetric
  Capillary Microchannel with Sudden Expansion},'' {\em Japanese Journal of
  Applied Physics}, vol.~47, no.~3, pp.~1683--1689, 2008.

\bibitem{Taher2018}
A.~Taher, B.~Jones, P.~Fiorini, and L.~Lagae, ``{Analytical, numerical and
  experimental study on capillary flow in a microchannel traversing a backward
  facing step},'' {\em International Journal of Multiphase Flow}, vol.~107,
  pp.~221--229, 2018.

\bibitem{Kaufman2017}
Y.~Kaufman, S.-Y. Chen, H.~Mishra, A.~M. Schrader, D.~W. Lee, S.~Das, S.~H.
  Donaldson, and J.~N. Israelachvili, ``{Simple-to-Apply Wetting Model to
  Predict Thermodynamically Stable and Metastable Contact Angles on
  Textured/Rough/Patterned Surfaces},'' {\em The Journal of Physical Chemistry
  C}, vol.~121, no.~10, pp.~5642--5656, 2017.

\bibitem{Panter2019}
J.~R. Panter, Y.~Gizaw, and H.~Kusumaatmaja, ``{Multifaceted design
  optimization for superomniphobic surfaces},'' {\em Science Advances}, vol.~5,
  no.~6, p.~eaav7328, 2019.

\bibitem{Ma2019}
B.~Ma, L.~Shan, B.~Dogruoz, and D.~Agonafer, ``{Evolution of Microdroplet
  Morphology Confined on Asymmetric Micropillar Structures},'' {\em Langmuir},
  vol.~35, no.~37, pp.~12264--12275, 2019.

\bibitem{Agonafer2018}
D.~D. Agonafer, H.~Lee, P.~A. Vasquez, Y.~Won, K.~W. Jung, S.~Lingamneni,
  B.~Ma, L.~Shan, S.~Shuai, Z.~Du, T.~Maitra, J.~W. Palko, and K.~E. Goodson,
  ``{Porous micropillar structures for retaining low surface tension
  liquids},'' {\em Journal of Colloid and Interface Science}, vol.~514,
  pp.~316--327, 2018.

\bibitem{DeGennes2004}
P.-G. de~Gennes, F.~Brochard-Wyart, and D.~Qu{\'{e}}r{\'{e}}, {\em {Capillarity
  and Wetting Phenomena}}.
\newblock New York, NY: Springer New York, 2004.

\bibitem{Moebius2012}
F.~Moebius and D.~Or, ``{Interfacial jumps and pressure bursts during fluid
  displacement in interacting irregular capillaries},'' {\em J. Colloid
  Interface Sci.}, vol.~377, pp.~406--415, 2012.

\bibitem{Rabbani2019}
H.~S. Rabbani and T.~D. Seers, ``{Inertia Controlled Capillary Pressure at the
  Juncture between Converging and Uniform Channels},'' {\em Sci. Rep.}, vol.~9,
  p.~13870, 2019.

\bibitem{Gibbs1906}
J.~W. Gibbs, {\em {The Scientific Papers of J. Willard Gibbs, Volume 1.
  Thermodynamics}}.
\newblock London: Longmans, Green and Co., 1906.

\bibitem{Blow2009}
M.~L. Blow, H.~Kusumaatmaja, and J.~M. Yeomans, ``{Imbibition through an array
  of triangular posts},'' {\em Journal of Physics: Condensed Matter}, vol.~21,
  no.~46, p.~464125, 2009.

\bibitem{Lee2008}
S.~Lee, J.-S. Park, and T.~R. Lee, ``{The Wettability of Fluoropolymer Surfaces
  : Influence of Surface Dipoles},'' {\em Langmuir}, vol.~24, no.~9,
  pp.~4817--4826, 2008.

\end{thebibliography}


\cleardoublepage

\setcounter{figure}{0}
\makeatletter 
\renewcommand{\thefigure}{S\@arabic\c@figure}
\makeatother

\begin{huge}
Supporting information
\end{huge}

\section*{Critical pressure asymmetry limits}

\begin{figure}[!ht]
\includegraphics[width=0.5\textwidth]{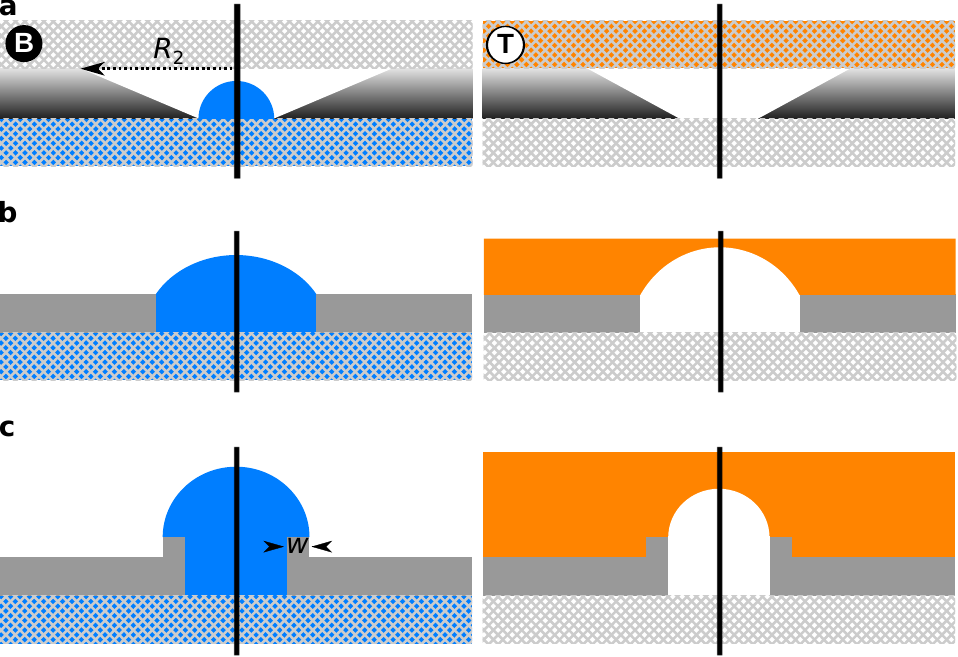}
\caption{Illustrations of the constructions used to calculate the significant critical pressure limits. (a) Critical pressure types for a fully enclosed pore as $R_2' \rightarrow \infty$. The maximum asymmetry here is $A=1$. (b) Critical pressure types for a half-enclosed cylindrical pore. The maximum asymmetry here is $A=\sqrt{2}$. (c) Maximum possible asymmetry on a structured pore. The maximum asymmetry is $A=2$.}
\label{fig:appb}
\end{figure}

In Fig. \ref{fig:appb}, we define the critical pressure asymmetry limits we use in the main text. The first highlighted asymmetry, $A=1$, is the maximum possible asymmetry for a doubly enclosed conical pore when $R_2 \rightarrow \infty$, illustrated in Fig. \ref{fig:appb}(a) for $R_2 \rightarrow \infty$. When $\theta_1 > \alpha + \pi/2$ and $L'>1$, but $\theta_1 + \alpha < \pi$ and $\theta_2 + \alpha < \pi$, then $\Delta P_c^{\rm{out}}(B)=1$ and $\Delta P_c^{\rm{in}}(T) \rightarrow 0$, leading to $A \rightarrow 1$.

The second highlighted asymmetry, $A=\sqrt{2}$, is the maximum possible asymmetry for a half-open cylindrical pore, illustrated in Fig. \ref{fig:appb}(b). Here, if $\theta_1 = \theta_2 = \pi/4$, then $\Delta P_c^{\rm{out}}=\sqrt{2}/2$ and $\Delta P_c^{\rm{in}}(T)=-\sqrt{2}/2$, leading to $A=\sqrt{2}$

The third highlighted asymmetry, $A=2$, is the maximum possible asymmetry for any pore geometry, illustrated in Fig. \ref{fig:appb}(c). Here, a lip structure is required on a half-open pore, with $\theta_1 = \theta_2 = 0$. As the lip width $w \rightarrow 0$, $\Delta P_c^{\rm{out}} \rightarrow 1$ and $\Delta P_c^{\rm{in}} \rightarrow -1$, leading to $A \rightarrow 2$. Such an intricately patterned pore design has been recently developed \cite{Agonafer2018}. The disadvantage of this geometry is that it required complex manufacturing procedures, and is not suitable for enclosed-pore applications.

\section*{Proof of $A>0$}
\begin{figure}[!ht]
\includegraphics[width=0.5\textwidth]{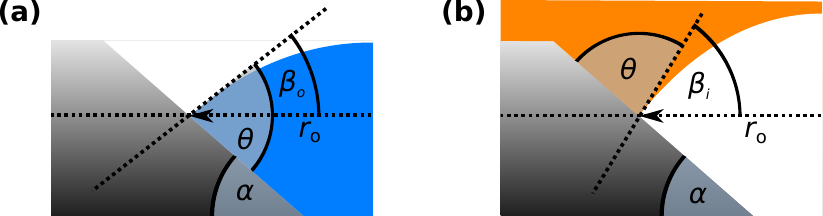}
\caption{Construction of the incoming (a) and outgoing (b) meniscus of a single pore for a fixed contact line radius $r_o$.}
\label{fig:beta}
\end{figure}

Here we prove that for $\alpha<\pi/2$, the asymmetry $A>0$. In Fig. \ref{fig:beta}(a), we introduce the angle $\beta_o$, which determines the sign and magnitude of the reduced outgoing pressure at a fixed contact line radius. We define $\beta_o$ with respect to the horizontal axis, such that $\beta_o = \theta - \alpha$. If $\beta_o >0$, the meniscus is convex, whereas if $\beta_o < 0$, the meniscus is concave. The larger the $|\beta_o|$, the larger the magnitude of the outgoing Laplace pressure. In Fig. \ref{fig:beta}(b), we introduce the angle $\beta_i$, which determines the sign and magnitude of the reduced incoming pressure at a chosen contact line radius $r_o$. We define $\beta_i$ with respect to the horizontal axis, such that $\beta_i = \theta + \alpha - \pi$. If $\beta_i >0$, the meniscus is convex, whereas if $\beta_i < 0$, the meniscus is concave. The larger the $|\beta_i|$, the larger the magnitude of the incoming Laplace pressure. When the contact line is at the same location in the pore for the outgoing and incoming interface, we ask whether it is possible for $\beta_o<\beta_i$. This is only possible if $\alpha>\pi/2$, which would require an an inverted pore. For $\alpha<\pi/2$ considered here, $\beta_o>\beta_i$ at all contact line radii. It therefore follows that $\Delta P_r^{\rm{out}}(r_o) > \Delta P_r^{\rm{in}}(r_o)$. Since $\Delta P_c^{\rm{out}}$ is the maximum of $\Delta P_r^{\rm{out}}(r)$, and $\Delta P_c^{\rm{in}}$ is the maximum of $\Delta P_r^{\rm{in}}(r)$ for $r \in [1, R_2]$, then it is necessarily true that for $\alpha<\pi/2$, $\Delta P_c^{\rm{out}} > \Delta P_c^{\rm{in}}$, so that $A>0$.

\end{document}